\documentclass[11pt,bezier]{article}
\usepackage{amsmath,amssymb,amsfonts,euscript}
\usepackage[usenames]{color}
\newpage

\newcommand{\be}{\begin{equation}}
\newcommand{\ee}{\end{equation}}
\newtheorem{theorem}{Theorem}[section]
\newtheorem{proof}{Proof}[section]

\newtheorem{remark}{Remark}[section]

\title{\bf\Large   Markov Switching  Component \textbf{GARCH} Model: Stability and Forecasting}
\vspace{-1cm}
\author
{N. Alemohammad, S. Rezakhah, S. H. Alizadeh \footnote{Faculty of Mathematics and Computer Science, Amirkabir University of Technology,
Tehran, Iran. Email: n-alemohammad@aut.ac.ir, rezakhah@aut.ac.ir, sasan.h.alizadeh@qiau.ac.ir}}
\begin{document}
\maketitle
\begin{abstract}
This paper introduces  an extension of   the  Markov switching \textbf{GARCH} model  where  the volatility in each state is a convex combination of two different  \textbf{GARCH} components with time varying weights. \textbf{This model has the dynamic behaviour to capture the variants of shocks.} The asymptotic behavior of the second moment is investigated and  an appropriate upper bound for it is evaluated. The estimation of the  parameters by using  the Bayesian method via Gibbs sampling algorithm is studied. Finally we illustrate the efficiency of the  model by simulation and \textbf{empirical analysis.} We show that this model provides a much better forecast of the volatility than the Markov switching GARCH model.
 \quad\\

{\it Keywords}:  \textbf{GARCH} models,  Markov process, Stability, Component \textbf{GARCH} models, Forecasting, Bayesian inference, Griddy Gibbs sampling.  \\ \quad \\
{\it Mathematics Subject Classification:} 60J10, 62M10, 62F15.

\end{abstract}

 \section{Introduction}
 \par In the past three  decades, there has been a growing interest in using non linear time series models in finance and economy.  For financial time series, the ARCH   and
GARCH models, introduced by Engle \cite{engle a} and Bollerslev \cite{bollerslev}, are surely the most popular classes of volatility models. Although these models have been applied extensively in the modeling of financial time series, the dynamic structure of volatility can not be captured passably by such models. For more consistent volatility modeling, the models by time varying parameters are introduced. One class of such models is that of smooth transition GARCH models  presented by Gonzalez-Rivera \cite{gonzalez-rivera}, Lubrano
 \cite{lubrano} (see also  Hagerud \cite{hagerud} and Medeiros and Veiga \cite{medeiros}). These models can be considered as a valuable tool for including the asymmetry properties to negative and positive or small and big shocks in financial time series. The  component GARCH models, introduced first by Ding and Granger \cite{ding},  are also a generalization of the constant parameter GARCH model. In the structure of the component GARCH model (\cite{ding}), two different GARCH components contribute to the overall conditional variance at time t. One component has the high volatility (integrated variance component) and the other one has the low volatility.  These models have been widely applied in modeling the financial time series (e.g. \cite{maheu} and \cite{engle c}). A generalization of the component GARCH model of Ding and Granger is the weighted GARCH model that is proposed by Bauwens and Storti \cite{bauwens b}.  In this model the weights of GARCH components  are the functions of lagged values of the conditional standard deviation or squared past observations. \\
 \par Another class is that of  Markov switching models.  These models are obtained by   Merging (G)ARCH model with a  Markov process, where each state of the Markov model allows a different (G)ARCH behavior. These models are  introduced by Cai \cite{cai} and Hamilton and Susmel \cite{hamilton}. This feature   extends the dynamic formulation of the model and potentially enables improving forecasts of the volatility \cite {abramson}. Gray \cite{gray}, Klaassen \cite{klaassen}, Haas, Mittnik and Paolella \cite{haas} proposed different variants of Markov-Switching GARCH models. See also further studies,  Abramson and Cohen \cite{abramson}, Alexander and Lazar \cite{alexander} and Bauwens et al. \cite{bauwens c}.

\par In this paper we consider a Markov switching model that the volatility of each state is a convex combination of two GARCH regimes with time varying coefficients which is in effect of the  previous observation. \textbf{This model has the potential to switch between several regimes with various volatilities  and also is able to model the time series with variants of shocks. The structure of the model makes a dynamic behavior in each regime to react differently to the species of shocks. For example in the high volatility regime, the model is able to have different responses to very high and high shocks and in low volatility regime different reactions to moderate and  low shocks.} We consider different weight functions for each state that allow volatility in each state  to react differently to the shocks of equal size. As using all past observations for forecasting could increase the complexity of the model, we reduce the volume of calculations  by proposing a dynamic programming algorithm. We derive necessary and sufficient conditions for stability and obtain an upper bound for the limit of the second moment by using the method of Abramson and Cohen \cite{abramson} and Medeiros \cite{medeiros}. For the estimation of the parameters, we use  the Bayesian inference via the Gibbs sampling. We  compare the performance of our model with the Markov switching GARCH model. The Markov switching component GARCH model can forecast the conditional variance much better  than MS-GARCH model.
\\
 \par The paper is organized as follows: in section 2 we introduce the  Markov switching component  GARCH model. Section 3 investigates the statistical properties of the model. Section 4 is devoted to the  estimation of the parameters of the model. Section 5 is dedicated to the analyzing of the efficiency of the proposed model through simulation and  the comparison of the forecast errors with the MS-GARCH model. \textbf{The empirical applications and discussion are developed in section 6.} Section 7 concludes.

\section{  Markov Switching Component GARCH Model}
The Markov switching component \textbf{GARCH} model, \textbf{MS-CGARCH}, for time series $\{y_t\}$ is defined as
 \begin{equation}\label{1}
 y_t=\varepsilon_{t}\sqrt{H_{t,Z_t}},\hspace{2cm}
\end{equation}
where $\{\varepsilon_{t}\}$ are iid  standard normal variables, $\{Z_t\}$ is an irreducible and aperiodic  Markov chain on finite state space $E=\{1,2,\cdots,K\}$
 with transition probability matrix
$\; P=||p_{ij}||_{K\times K},\;$
where  $\, p_{ij}=p(Z_t=j|Z_{t-1}=i),\; i,j \in \{1,\cdots,K\}$,
and stationary probability measure
$\,
\pi=(\pi_1,\cdots,\pi_K)^{\prime}.$  Also given that $Z_t=j$, $H_{t,j}$ (the conditional variance in regime j) is driven by
\begin{equation}\label{2}
H_{t,j}=w_{t,j}h_{1,t,j}+(1-w_{t,j})h_{2,t,j},
\end{equation}
where
\begin{align}\label{3}
h_{1,t,j}= &a_{0j}+a_{1j}y^2_{t-1}+a_{2j}H_{t-1,j},\\
h_{2,t,j}=&b_{0j}+b_{1j}y^2_{t-1}+b_{2j}H_{t-1,j},
 \end{align}
 and each of the weights ($w_{t,j}$) is a function of the past observation as
\begin{equation}\label{4}
w_{t,j}=\frac{1-\exp(-\gamma_j|y_{t-1}|)}{1+\exp(-\gamma_j|y_{t-1}|)}\ \ \ \  \gamma_j>0,\ \ \ \
\end{equation}

 \noindent   which is  bounded , $0<w_{t,j}<1$. The parameter $\gamma_j$ is called the slope parameter, that
  explains  the speed of transition from one component to the other one: the higher $\gamma_j$, the faster the transition. $H_{t-1,j}$ in (\ref{2}) is the conditional variance of state m at time $t-1$, that is a combination of conditional variances of both components at the state j. Since   $\gamma_j>0$, when the absolute value of $y_{t-1}$ increases, the impact of $h_{1,t,j}$  increases and consequently the effect of $h_{2,t,j}$ decreases and vice versa. \textbf{Another good feature of our model is that it overcomes the problem of path dependency \footnote{Path dependency happens when the volatility of each regime at time t depends on the entire sequence of past regimes because of the recursive property of GARCH processes.} (that is common in some kinds of MS-GARCH processes). \\If $w_{t,j}$ becomes a constant value (for example when $\gamma_j$ tending to zero or infinity), the MS-CGARCH model will be a  MS-GARCH model. In the case of single regime, if $a_{2.}=b_{2.}$, our model is the generalization of the smooth transition GARCH model that is introduced by  Lubrano \cite{lubrano}.}

 \par It is assumed that $\{\varepsilon_{t}\}$ and  $\{Z_t\}$ are independent.  Sufficient conditions to guarantee strictly positive conditional variance  are
 $a_{0j},b_{0j}$ to be positive and $a_{1j},a_{2j},b_{1j}\\,b_{2j}$
being nonnegative.
\\
\par Let  $\mathcal{I}_{t}$  be the observation set up to time t. The conditional density function of $y_t$ given past observations is obtained as follows:\\
\begin{align}\label{7}
f(y_t|\mathcal{I}_{t-1})= &\sum_{j=1}^{K}{f(y_t,Z_t=j|\mathcal{I}_{t-1})}
\nonumber\\
&=\sum_{j=1}^{K}{p(Z_t=j|\mathcal{I}_{t-1})f(y_t|\mathcal{I}_{t-1},Z_t=j)}
\nonumber\\
&=\sum_{j=1}^{K}{\alpha_j^{(t)}\phi(\frac{y_t}{\sqrt{H_{t,j}}})}
\end{align}
in which $\alpha_j^{(t)}=p(Z_t=j|\mathcal{I}_{t-1})$ (that is  obtained in next section), and $\phi(.)$ is the probability density function of the standard normal distribution.

\section{Statistical Properties of the model}
\par In this section,  the statistical properties of the  \textbf{MS-CGARCH} model are investigated and   the conditional variance of the process is obtained. We show that the  model, under some conditions on coefficients and transition  probabilities , is asymptotically stable in the second moment. An appropriate
upper bound for the limiting value of the second moment is obtained.
\\
\subsection{Forecasting}
\par The forecasting volatility (conditional variance) of \textbf{MS-CGARCH} model is given by
\begin{equation}\label{8}
Var(Y_t|\mathcal{I}_{t-1}) =\sum_{j=1}^{K}{\alpha_j^{(t)}H_{t,j}}=\sum_{j=1}^{K}{\alpha_j^{(t)}(w_{t,j}h_{1,t,j}+(1-w_{t,j})h_{2,t,j})}.
\end{equation}
 This relation shows that the conditional variance of this model is affected by the changes in states, the volatility of components and the weight functions in each state. \\
At each time $t$, $\alpha_j^{(t)}$ (in equation (\ref{7}), (\ref{8})) can be obtained from a dynamic programming method based on forward recursion algorithm, proposed in remark (\ref{9}).
{{{\remark{{\label{9}}
The value of $\alpha_j^{(t)}$  is obtained recursively by
\begin{equation}
\alpha_j^{(t)}=\frac{\sum_{m=1}^{K}{f(y_{t-1}|Z_{t-1}=m,\mathcal{I}_{t-2})p(Z_{t-1}=m|\mathcal{I}_{t-2})}p_{m,j}}{\sum_{m=1}^{K}{f(y_{t-1}|Z_{t-1}=m,\mathcal{I}_{t-2})p(Z_{t-1}=m|\mathcal{I}_{t-2})}}.
\end{equation}
}}
{{\proof As the hidden variables $\{Z_{t}\}_{t\geq 1}$ have Markov structure in the MS-CGARCH model, so
\begin{align}
\alpha_j^{(t)}= &p(Z_t=j|\mathcal{I}_{t-1})=\sum_{m=1}^{K}{P(Z_t=j,Z_{t-1}=m|\mathcal{I}_{t-1})}
\nonumber\\
&=\sum_{m=1}^{K}{p(Z_t=j|Z_{t-1}=m,\mathcal{I}_{t-1})p(Z_{t-1}=m|\mathcal{I}_{t-1})}
\nonumber\\
&=\sum_{m=1}^{K}{p(Z_t=j|Z_{t-1}=m)p(Z_{t-1}=m|\mathcal{I}_{t-1})}
\nonumber\\
&=\frac{\sum_{m=1}^{K}{f(\mathcal{I}_{t-1},Z_{t-1}=m)p_{m,j}}}{\sum_{m=1}^{K}{f(\mathcal{I}_{t-1},Z_{t-1}=m)}}
\nonumber\\
&=\frac{\sum_{m=1}^{K}{f(y_{t-1}|Z_{t-1}=m,\mathcal{I}_{t-2})p(Z_{t-1}=m|\mathcal{I}_{t-2})}p_{m,j}}{\sum_{m=1}^{K}{f(y_{t-1}|Z_{t-1}=m,\mathcal{I}_{t-2})p(Z_{t-1}=m|\mathcal{I}_{t-2})}},
\end{align}
where
$$f(y_{t-1}|Z_{t-1}=m,\mathcal{I}_{t-2})=\phi(\frac{y_{t-1}}{\sqrt{H_{t-1,m}}}).\hspace{4cm}$$}}
\subsection{Stability}
\par In this subsection, we investigate the stability of the second moment of  MS-CGARCH model. Indeed we are looking for an upper bound for the second moment of our model.
The second moment  of the model can be calculated as:
$$E(y^2_t)=E(H_{t,Z_t})=E_{Z_t}[E_{t-1}(H_{t,Z_t}|z_t)]\hspace{6cm}$$

\begin{equation}\label{10}
=\sum_{z_t=1}^{K}{\pi_{z_t}E_{t-1}(H_{t,Z_t}|z_t)}.\hspace{2.5cm}
\end{equation}

$E_{t}(\cdot)$  denotes the expectation with respect to  the information up to time t. Also
for summarization, we shall use $E(\cdot |z_t)$ and $p(\cdot |z_t)$ to represent $E(\cdot |Z_t=z_t)$ and $P(\cdot|Z_t=z_t)$, respectively, where $z_t$ is the realization of the state at time t. We investigate the conditional variance under the  chain state, $m$, as follows:
\begin{align}\label{11}
E_{t-1}(H_{t,m}|z_t)&\nonumber
=E_{t-1}[w_{t,m}(a_{0m}+a_{1m}y^2_{t-1}+a_{2m}H_{t-1,m})|z_t]\hspace{5cm}\\
\nonumber
&\hspace{2cm}+E_{t-1}[(1-w_{t,m})(b_{0m}+b_{1m}y^2_{t-1}+b_{2m}H_{t-1,m})|z_t]\\
\nonumber
&=b_{0m}+\underbrace{b_{1m}E_{t-1}[y^2_{t-1}|z_t]}_{I}+\underbrace{(a_{0m}-b_{0m})E_{t-1}[w_{t,m}|z_t]}_{II}+\underbrace{b_{2m}E_{t-1}(H_{t-1,m}|z_t)}_{III}
\nonumber \\
\hspace{2.5cm}&+\underbrace{(a_{1m}-b_{1m})E_{t-1}[w_{t,m}y^2_{t-1}|z_t]}_{IV}+
\underbrace{(a_{2m}-b_{2m})E_{t-1}(w_{t,m}H_{t-1,m}|z_t)}_{V}.\\
\nonumber
\end{align}
The relation (II) in (\ref{11}) can be interpreted  as follows:

 $$E_{t-1}[y^2_{t-1}|z_t]=\sum_{z_{t-1}=1}^{K}
 {\int_{S_{\mathcal{I}_{t-1}}}{y^2_{t-1}p(\mathcal{I}_{t-1}|z_t,z_{t-1})p(z_{t-1}|z_t)d\mathcal{I}_{t-1}}}$$
 \begin{equation}\label{12}
 =\sum_{z_{t-1}=1}^{K}{p(z_{t-1}|z_t)E_{t-1}[y^2_{t-1}|z_{t-1},z_t]},
 \end{equation}
 where $S_{\mathcal{I}_{t-1}}$ is the support of $\mathcal{I}_{t-1}=(y_1,\cdots,y_{t-1})$. Since the expected value of $y^2_{t-1}$ is independent of  any future  state, so
 \begin{equation}\label{13}
 E_{t-1}[y^2_{t-1}|z_{t-1},z_t]=E_{t-1}[y^2_{t-1}|z_{t-1}].
\end{equation}
\par Also using the tower property of the conditional expectation, $E[E(Y|X,Z)|X]=E(Y|X)$ [see Grimmett and Stirzaker (2001, p. 69)], we have
$$E_{t-1}[y^2_{t-1}|z_{t-1}]
=E_{t-2}[E_{t-1}(y^2_{t-1}|\mathcal{I}_{t-2},z_{t-1})|z_{t-1}]\hspace{9.3cm}$$
  \begin{equation}\label{14}
 \hspace{2.6cm}=E_{t-2}[H_{t-1,Z_{t-1}}|z_{t-1}].\hspace{5cm}
  \end{equation}
 \par The calculation of $E_{t-1}[w_{t,m}|z_t]$, $E_{t-1}[w_{t,m}y^2_{t-1}|z_t]$  and $E_{t-1}(w_{t,m}H_{t-1,m}|z_t)$ is a problem that can not be easily done, for this reason we will try to find an upper bound for them.\\
  \\
\emph{Upper bound to II.} As $0<w_{t,m}<1$, so an upper bound for the relation II in (\ref{11}) is obtained by
 \begin{equation}\label{15}
 (a_{0m}-b_{0m})E_{t-1}[w_{t,m}|z_t]\leq |a_{0m}-b_{0m}|< \infty.
  \end{equation}
  \textbf{The relation (III) can be specified as
  \begin{equation*}
  b_{2m}E_{t-1}(H_{t-1,m}|z_t)= b_{2m}{\int_{S_{\mathcal{I}_{t-1}}}{H_{t-1,m}p(\mathcal{I}_{t-1}|z_t)d\mathcal{I}_{t-1}}}
    \end{equation*}
 \begin{equation}
    = b_{2,m}\sum_{z_{t-1}=1}^{K}
 {p(z_{t-1}|z_t)E_{t-2}(H_{t-1,m}|z_{t-1})}.
 \end{equation}}
  \emph{Upper bound to IV.}  Let  $0<M<\infty$ be a constant, so
\begin{align}
E_{t-1}[w_{t,z_t}y^2_{t-1}|z_t]= &E_{t-1}[w_{t,z_t}y^2_{t-1}I_{|y_{t-1}|<M}|z_t]
\nonumber\\
&+E_{t-1}[w_{t,z_t}y^2_{t-1}I_{|y_{t-1}|\geq M}|z_t]
\nonumber
\end{align}
in which
\begin{displaymath}
I_{x<a}=\left\{ \begin{array}{ll}
1 & \textrm{if  $x<a$}\\
0 & \textrm{otherwise.}
\end{array} \right.
\end{displaymath}
As by (\ref{4}),  $0<w_{t,z_t}<1$ and so
\begin{equation*}
E_{t-1}[w_{t,z_t}y^2_{t-1}|z_t]\leq M^2+E_{t-1}[w_{t,z_t}y^2_{t-1}I_{|y_{t-1}|\geq M}|z_t],
\end{equation*}
also
\begin{align}
E_{t-1}[w_{t,z_t}y^2_{t-1}I_{|y_{t-1}|\geq M}|z_t]=&\int_{S_{\mathcal{I}_{t-2}},y_{t-1}\leq -M}{y^2_{t-1}[w_{t,z_t}]p(\mathcal{I}_{t-1}|z_t)d\mathcal{I}_{t-1}}
\nonumber\\
&+\int_{S_{\mathcal{I}_{t-2}},y_{t-1}\geq M}{y^2_{t-1}[w_{t,z_t}]p(\mathcal{I}_{t-1}|z_t)d\mathcal{I}_{t-1}},
\nonumber
\end{align}
by (\ref{4}),
\textbf{\begin{equation}
\lim_{y_{t-1}\rightarrow +\infty}w_{t,z_t}=1, \ \ \ \ \and\ \ \ \ \lim_{y_{t-1}\rightarrow -\infty}w_{t,z_t}=1,
\end{equation}
therefore according to the definition of limit at infinity, for a small number $\delta>0$, there will exist a finite constant $M>0$ such that if $y_{t-1}\geq M$, $|w_{t,z_t}-1|\leq \delta$ and if $y_{t-1}\leq -M$, $|w_{t,z_t}-1|\leq \delta$. Hence
\begin{align}
E_{t-1}[w_{t,z_t}y^2_{t-1}I_{|y_{t-1}|\geq M}|z_t] \leq \ & (\delta+1) \int_{S_{\mathcal{I}_{t-2}},y_{t-1}\leq -M}{y^2_{t-1}p(\mathcal{I}_{t-1}|z_t)d\mathcal{I}_{t-1}}
\nonumber\\
&+ (\delta+1) \int_{S_{\mathcal{I}_{t-2}},y_{t-1}\geq M}{y^2_{t-1}p(\mathcal{I}_{t-1}|z_t)d\mathcal{I}_{t-1}}.
\nonumber
\end{align}}
Since the distribution of the $\{\varepsilon_{t}\}$ is symmetric, then
\begin{align}
(\delta+1)\int_{S_{\mathcal{I}_{t-2}},y_{t-1}\leq -M}{y^2_{t-1} p(\mathcal{I}_{t-1}|z_t)d\mathcal{I}_{t-1}}\leq & (\delta+1) \int_{S_{\mathcal{I}_{t-2}},-\infty<y_{t-1}<0}{y^2_{t-1}p(\mathcal{I}_{t-1}|z_t)d\mathcal{I}_{t-1}}
\nonumber\\
&=(\delta+1)\frac{ E_{t-1}[y^2_{t-1}|z_t]}{2}
\nonumber
\end{align}
\\
and
\begin{align}
(\delta+1)\int_{S_{\mathcal{I}_{t-2}},y_{t-1}\geq M}{y^2_{t-1}p(\mathcal{I}_{t-1}|z_t)d\mathcal{I}_{t-1}}\leq & (\delta+1)\int_{S_{\mathcal{I}_{t-2}},0<y_{t-1}<\infty}{y^2_{t-1} p(\mathcal{I}_{t-1}|z_t)d\mathcal{I}_{t-1}}
 \nonumber\\
 &=(\delta+1)\frac{ E_{t-1}[y^2_{t-1}|z_t]}{2}.
 \nonumber
\end{align}
\\
Therefor
\begin{equation*}
(a_{1m}-b_{1m})E_{t-1}[w_{t,z_t}y^2_{t-1}|z_t]\leq  |a_{1m}-b_{1m}|(M^2+(\delta+1) E_{t-1}[y^2_{t-1}|z_t]).
\end{equation*}
 \textbf{\emph{Upper bound to V.} Since $0<w_{t,m}<1$, so
  \begin{equation}
  (a_{2m}-b_{2m})E_{t-1}(w_{t,m}H_{t-1,m}|z_t)\leq|a_{2m}-b_{2m}|E_{t-1}(H_{t-1,m}|z_t).
   \end{equation}}
 \\
  By replacing the obtained  upper bounds and relations (\ref{12})-(\ref{14}) in (\ref{11}),  the upper bound for $E_{t-1}(H_{t,Z_t}|z_t)$ is acquired as:
\begin{align}
E_{t-1}(H_{t,m}|z_t)&\leq a_{0m}+|a_{1m}-b_{1m}|M^2
\nonumber\\
&+\sum_{z_{t-1}=1}^{K}{{[b_{1m}+|a_{1m}-b_{1m}|(\delta+1)]
p(z_{t-1}|z_t)}E_{t-2}[H_{t-1,Z_{t-1}}|z_{t-1}]}
\nonumber\\
+&\sum_{z_{t-1}=1}^{K}{{a_{2m}
p(z_{t-1}|z_t)}E_{t-2}[H_{t-1,m}|z_{t-1}]},
\end{align}
in which by Bayes' rule
$$p(z_{t-1}|z_t)=\frac{\pi_{z_{t-1}}}{\pi_{z_t}}\{P_{z_{t-1}z_t }\},$$
where $P$ is the transition probability matrix.
\textbf{Let
 \begin{equation}\label{16}
 {\bf{\Omega}}=[a_{01}+|a_{11}-b_{11}|M^2,\cdots,\\a_{0K}+|a_{1K}-b_{1K}|M^2]^{\prime},
 \end{equation}
  be a vector with K component, ${\bf{C}}$ denotes  a $K^2$-by-$K^2$ block matrix as \\
  \begin{equation}\label{17}
  {\bf{C}}=\left(
             \begin{array}{cccc}
              {\bf{C}_{11}}& {\bf{C}_{21}}& \cdots & {\bf{C}_{K1}} \\
               {\bf{C}_{12}} & {\bf{C}_{22}} & \cdots & {\bf{C}_{K2}} \\
               \vdots &  &  & \vdots\\
               {\bf{C}_{1K}} & {\bf{C}_{2K}} & \cdots & {\bf{C}_{KK}} \\
             \end{array}
           \right)
  \end{equation}
  with each block given by
 \begin{equation}\label{18}
 {\bf{C}_{jk}}=p(Z_{t-1}=j|Z_t=k)( {\bf{u}}{\bf{e}}^{\prime}_j+{\bf{v}}),\ \ \ \ \ \ \quad\ j,k=1,\cdots,K,
 \end{equation}
 where $ {\bf{u}}=[b_{11}+(\delta+1)|a_{11}-b_{11}|,\cdots,b_{1K}+(\delta+1)|a_{1K}-b_{1K}|]^{\prime}$,   ${\bf{e}}_j$ is a K-by-1 vector of all zeros, except its jth element, which is one, and ${\bf{v}}$ is a diagonal K-by-K matrix with elements $[a_{21},\cdots,a_{2K}]$ on its diagonal.\\
Let $A_t(j,k)=E_{t-1}[H_{t,j}|Z_{t}=k]$, $\textbf{A}_t=[A_t(1,1),A_t(2,1),\cdots,A_t(K,1),A_t(1,2),\cdots,\\A_t(K,K)]$ be a $K^2$-by-1  vector and consider $\dot{ {\bf{\Omega}}}=( {\bf{\Omega}}^{\prime},\cdots, {\bf{\Omega}}^{\prime})^{\prime}$ be a  vector that is made of K vector  ${\bf{\Omega}}$.}

Hence by  (\ref{16})-(\ref{18}) we have the following recursive inequality vector form for $\textbf{A}_t$, as
\begin{equation}\label{19}
\textbf{A}_t\leq \dot{ {\bf{\Omega}}} + {\bf{C}}\textbf{A}_{t-1},\ \ \ t\geq 0.
\end{equation}
with some initial conditions $\textbf{A}_{-1}.$\\
\textbf{Let $\Pi=[\pi_1{\bf{e}}^{\prime}_1,\cdots,\pi_K{\bf{e}}^{\prime}_K]$ }and consider  $\rho(A)$ denotes  the spectral radius of a matrix A, then we have the following theorem for the stationarity condition of the \textbf{MS-CGARCH} model.
{\theorem Let $\{Y_t\}_{t=0}^{\infty}$ follows the MS-CGARCH  model, defined by (\ref{1})-(\ref{4}),
 the process is asymptotically stable in  variance  and $\lim_{t\rightarrow\infty}E(Y^2_t)\leq{\bf{\Pi^\prime(I-{\bf{C}})^{-1}\dot{{\bf{\Omega}}}}}$, if and only if $\rho({\bf{C}})
<1.$}
{\proof \cite{abramson}, By recursive inequality (\ref{19}),
\begin{equation}\label{20}
\textbf{A}_t\leq \dot{{\bf{\Omega}}}\sum_{i=0}^{t-1}{{\bf{C}}^i}+{\bf{C}}^{t}\textbf{A}_{0}:=\textbf{B}_{t}.\hspace{8.5cm}
\end{equation}
By the matrix convergence theorem \cite{lancaster}, a necessary and sufficient condition for the convergence of $\textbf{B}_{t}$ where $t\rightarrow\infty$
is $\rho({\bf{C}})
<1$ (  the value of $\delta$ can be considered  small enough  to be negligible). Under this condition, ${\bf{C}}^t$ converges to zero as t goes to infinity and $\sum_{i=0}^{t-1}{{\bf{C}}^i}$ converges to $(I-{\bf{C}})^{-1}$
provided that matrix $(I-{\bf{C}})$ is invertible.   So if $\rho({\bf{C}})
<1$,
$$\lim_{t\rightarrow\infty}\textbf{A}_t\leq(I-{\bf{C}})^{-1}\dot{{\bf{\Omega}}}.\hspace{8.5cm}$$
  By (\ref{10})
the upper bound for the asymptotic behavior of  unconditional variance is given by
$$lim_{t\rightarrow\infty}E(y^2_t)\leq{\bf{\Pi^\prime(I-C)^{-1}}}\dot{{\bf{\Omega}}}.\hspace{8cm}$$}}}
\textbf{If $\rho({\bf{C}})\geq 1$, $\textbf{B}_{t}$ goes to infinity with the growth of the time and it can not be possible to find an upper bound for the unconditional second moment of the model.}

\section{Estimation}
In this section we describe the estimation of the parameters of the MS-CGARCH model. We consider  Bayesian MCMC method using Gibbs algorithm  by following methods of sampling of a hidden Markov process (\cite{chib} and \cite{kaufman}), MS-GARCH model and weighted  GARCH model (\cite{bauwens b} and \cite{bauwens c} ) for  estimation of the parameters.\\
\par Let $Y_t=(y_1,\cdots,y_t)$ and $Z_t=(z_1,\cdots,z_t)$. For the case of two states,  the transition probabilities are $\eta=(\eta_{11},\eta_{12},\eta_{21},\eta_{22})$ and  the parameters of the model are $\theta=(\theta_1,\theta_2)$, where $\theta_k=(a_{0k},b_{0k},a_{1k},b_{1k},a_{2k},b_{2k},\gamma_k)$ for $k=1,2$.
\par The purpose of Bayesian inference is to simulate from the distributions of the parameters and the state variables given the observations. As $Z=(z_1,\cdots,z_T)$ and $Y=(y_1,\cdots,y_T)$, the posterior density of our model is:
\begin{equation}\label{21}
p(\theta,\eta,Z|Y)\propto p(\theta,\eta)p(Z|\theta,\eta)f(Y|\theta,\eta,Z),
\end{equation}
in which $p(\theta,\eta)$ is the prior of the parameters. The conditional probability mass function of $Z$ given the $(\theta,\eta)$ is independent of $\theta$, so
\begin{align}
p(Z|\theta,\eta)=& p(Z|\eta_{11},\eta_{22})
\nonumber\\
&=\prod_{t=1}^{T}{p(z_{t+1}|z_t,\eta_{11},\eta_{22})}
\nonumber\\
&=p_{11}^{n_{11}}(1-p_{11})^{n_{12}}p_{22}^{n_{22}}(1-p_{22})^{n_{21}},
\end{align}
where $n_{ij}=\#\{z_t=j|z_{t-1}=i\}$.
 The conditional density function of $Y$ given the realization of $Z$ and the parameters is factorized in the following way:
\begin{equation}
f(Y|\eta,\theta,Z)=\prod_{t=1}^{T}{f(y_t|\theta,z_t=k,Y_{t-1})},\ \ \ k=1,2,
\end{equation}
where the one step ahead of the predictive densities are:
\begin{equation}\label{22}
f(y_t|\theta,z_t=k,Y_{t-1})=\frac{1}{\sqrt{ 2\pi H_{t,k}}}\exp(-\frac{y^2_t}{H_{t,k}}).
\end{equation}
\par Since the posterior density (\ref{21}) is not standard we can not sample it in a straightforward manner. Gibbs sampling of Gelfand and Smith \cite{gelfand} is a repetitive algorithm to sample consecutively from the posterior distribution. Under regularity conditions, the simulated distribution converges to the posterior distribution, (see e.g Robert and  Casella \cite{robert}). The blocks of parameters are $\theta$, $\eta$ and the realizations of $Z$.\\
A brief description of the Gibbs algorithm:  Let use the superscript $(r)$ on $Z, \, \theta$ and $\eta$ to denote the estimators of
 $Z$,   $\eta$, and $\theta$ at the r-th iteration of the algorithm. Each iteration of the algorithm consists of three steps:\\
(i) Drawing an estimator  random sample of the state variable $Z^{(r)}$  given  $,\eta^{(r-1)},\ \theta^{(r-1)}$.\\
(ii) Drawing a random sample of the transition probabilities $\eta^{(r)}$ given $Z^{(r)}$.\\
(iii) Drawing a random sample of the $\theta^{(r)}$ given $Z^{(r)}$ and $\eta^{(r)}$.\\

These steps are repeated until the convergency is obtained. In what follows sampling of each block is explained. \\
\subsection{Sampling $z_t$}
The purpose of this step is to obtain the sample of  $p(z_t|\eta,\theta, Y_t)$ that is performed by Chib\cite{chib}, (see also \cite{kaufman}).  Suppose $p(z_1|\eta,\theta, Y_0,)$ be the stationary distribution of the chain,\\
\begin{equation}
p(z_t|\eta,\theta, Y_t)\propto f(y_t|\theta,z_t=k,Y_{t-1})p(z_t|\eta,\theta, Y_{t-1}),
\end{equation}
where the predictive density $ f(y_t|\theta,z_t=k,Y_{t-1})$ is calculated by the relation (\ref{22}) and by the law of total probability $p(z_t|\eta,\theta, Y_{t-1})$ is given by:\\
\begin{equation}
p(z_t|\eta,\theta, Y_{t-1})=\sum_{z_{t-1}=1}^{K}{p(z_{t-1}|\eta,\theta, Y_{t-1})\eta_{z_{t-1}z_t}}.
\end{equation}

Given the filter probabilities ($p(z_t|\eta,\theta, Y_t)$), we run a backward algorithm, starting from $t=T$ that $z_T$ is derived from $p(z_T|\eta,\theta,Y)$. For $t=T-1,\cdots,0$ the sample is derived from $p(z_t|z_{t+1},\cdots,z_T,\theta,\eta,Y)$,which is obtained by
\begin{equation*}
p(z_t|z_{t+1},\cdots,z_T,\theta,\eta,Y)\propto p(z_t|\eta,\theta,Y_t)\eta_{z_t,z_{t+1}}.
\end{equation*}
To derive  $z_t$ from $p(z_t|\cdot)=p_{z_t}$ is by sampling from the conditional probabilities (for example) $q_{1}=p(Z_t=1|Z_t\geq 1,.)$ which are given by
$$p(Z_t=1|Z_t\geq 1,.)=\frac{p_1}{\sum_{l=1}^{2}{p_l}}.$$
After generating a uniform (0,1) number $U$, if $U\leq q_1$ then $z_t=1$, otherwise $z_t=2$.
\subsection{Sampling $\eta$}
This stage is devoted to sample $\eta=(\eta_{11},\eta_{22})$ from the posterior probability $p(\eta|\theta, Y_t, Z_t)$ that is independent of $Y_t, \theta$. We consider independent beta prior density for each of   $\eta_{11}$ and $\eta_{22}$.    For example,
$$p(\eta_{11}|Z_t)\propto p(\eta_{11})p(Z_t|\eta_{11})=\eta_{11}^{c_{11}+n_{11}-1}(1-\eta_{11})^{c_{12}+n_{12}-1},$$
where $c_{11}$ and $c_{12}$ are the parameters of Beta prior, $n_{ij}$ is the number of transition from $z_{t-1}=i$ to $z_t=j$.  In the same way  the sample of $\eta_{22}$ is obtained.
\subsection{Sampling $\theta$}
The posterior density of $\theta$ given the prior $p(\theta)$ is given by:
\begin{equation}\label{23}
p(\theta|Y, Z, \eta)\propto p(\theta) \prod_{t=1}^{T}{f(y_t|\theta,z_t=k,Y_{t-1})}= p(\theta)\prod_{t=1}^{T}{\frac{1}{\sqrt{ 2\pi H_{t,k}}}\exp(-\frac{y^2_t}{H_{t,k}})},
\end{equation}
which is independent of $\eta$.
Since the conditional distribution of $\theta$ does not have a closed-form (because  for example $p(a_{0k}|Y_t,Z_t,\theta_{-a_{0k}})$, in which $\theta_{-a_{0k}}$ is the parameter vector without $a_{0k},$ contains $H_{t,k}$, which is also a function of $a_{0k}$. Therefor it can not be a normal density.) using the Gibbs sampling in this situation may be complicated. The Griddy Gibbs algorithm, that introduced by Ritter and Tanner (1992), can be a solution of this problem. This method is very applicable in researches (for example \cite{bauwens a} , \cite{bauwens b} and \cite{bauwens c}). \\
\par Given samples at iteration $r$ the Griddy Gibbs at iteration $r+1$ proceeds as follows:\\
\\
1. Select a grid of points, such as $a_{0i}^1,a_{0i}^2,\cdots,a_{0i}^G$. Using (\ref{23}), evaluate the conditional posterior density function $k(a_{0i|Z_t,Y_t,\theta_{-a_{0i}}})$ over the grid points to obtain the vector $G_k=(k_1,\cdots,k_G)$.\\
2. By a deterministic integration rule using the G points, compute
$G_{\Phi}=(0,\Phi_2,\cdots,\Phi_G)$
with
\begin{equation}
\Phi_j=\int_{a_{0i}^1}^{a_{0i}^j}{k(a_{01}|\theta_{-a_{0i}}^{(r)},Z^{(r)}_t,Y_t)da_{0i}},\ \ \ i=2,\cdots,G.
\end{equation}
3. Simulate $u\sim U(0,\Phi_G)$ and invert $\Phi(a_{0i}|\theta_{-a_{0i}}^{(r)},Z^{(r)}_t,Y_t)$ by numerical interpolation to obtain a sample $a_{0i}^{(r+1)}$ from $a_{0i}|\theta_{-a_{0i}}^{(r)},Z^{(r)}_t,Y_t$.\\
4. Repeat steps 1-3 for other parameters.
\\
\par For the prior densities of all elements of $\theta$, it can be can considered  independent uniform densities over the finite intervals.

\section{Simulation Results}
In this section we provide some simulation results of \textbf{MS-CGARCH} model defined by equations (\ref{1})-(\ref{4}) for two states. We  simulate 300 sample from the following MS-CGARCH model:
\begin{equation}
 y_t=\varepsilon_{t}\sqrt{H_{Z_t,t}},\hspace{10cm}
 \end{equation}
where $\{\varepsilon_{t}\}$ is an iid sequence of standard normal variables, $\{Z_t\}$ is a Markov chain on finite state space $E=\{1,2\}$
 with transition probability matrix
$$P=\left(
                  \begin{array}{cc}
                    .85 & .15   \\
                    .05 & .95  \\

                  \end{array}
                \right),$$
and
\begin{align}
H_{1,t}=&\frac{1-\exp(-2|y_{t-1}|)}{1+\exp(-2|y_{t-1}|)}(2.2+.75y^2_{t-1}+.15H_{1,t-1})+
\nonumber\\
&\hspace{4cm}[1-\frac{1-\exp(-2|y_{t-1}|)}{1+\exp(-2|y_{t-1}|)}](.7+.3y^2_{t-1}+.2H_{1,t-1}),
\nonumber\\
H_{2,t}=&\frac{1-\exp(-.5|y_{t-1}|)}{1+\exp(-.5|y_{t-1}|)}(.4+.15y^2_{t-1}+.1H_{2,t-1})+
\nonumber\\
&\hspace{4cm}[1-\frac{1-\exp(-.5|y_{t-1}|)}{1+\exp(-.5|y_{t-1}|)}](.2+.1y^2_{t-1}+.2H_{2,t-1}).
\end{align}
\par The first state implies a higher conditional variance than the second one and in each state, the first component has the higher volatility than the other component.
\label{fig1}
\input{epsf}
\epsfysize=2.5in
 \begin{figure}
\centerline{\epsffile{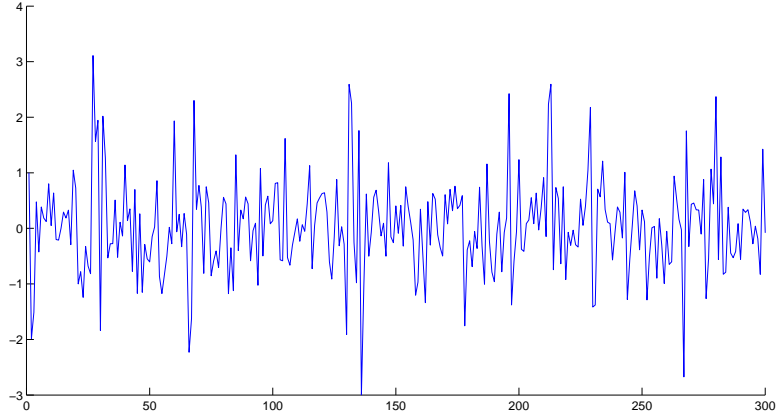}}
\vspace{-.2in}
\caption{\scriptsize Simulated time series of MS-CGARCH model.}
\end{figure}

\par In Table 1, we report summery statistics for  simulated data and  figure 1 shows  the  plot of the simulated time  series. \\
 \par Using the Bayesian inference, we estimate the parameters of the \textbf{MS-CGARCH} model.  The prior density of each parameter is assumed to be uniform restricted over a finite interval (except for $\eta_{11}$ and $\eta_{22}$, since they are drawn from the beta distribution). Table 2 demonstrates the performance of the estimation methods. The results of this table show that the  standard deviation are small enough in most cases. \\

\label{fig2}
\input{epsf}
\epsfysize=3.3in
 \begin{figure}
\centerline{\epsffile{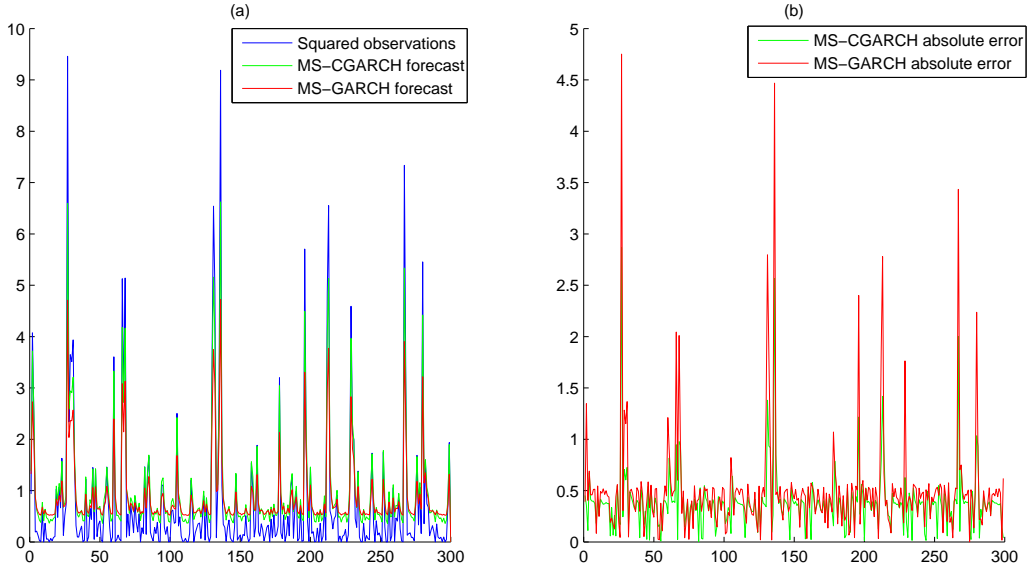}}
\vspace{-.2in}
\caption{\scriptsize (a): Squared observations of the simulated time series (blue), forecast by MS-GARCH (red) and forecast by MS-CGARCH (green). (b): Absolute forecast error of squared simulated time series  in the MS-GARCH (red) and in the MS-CGARCH (green).}
\end{figure}
 \par For clarifying  the performance of \textbf{MS-CGARCH} model toward MS-GARCH model, We  compare the forecasting volatility  ($E(Y^2_t|\mathcal{F}_{t-1})$) of each model with the squared observations. Figure 2 shows that the forecasting volatility of MS-CGARCH is much better than MS-GARCH model and the absolute forecast  error (the difference between the forecasting volatility and the squared observations)  of our model is often smaller than the MS-GARCH model. \textbf{The root of mean squared error of the MS-GARCH and MS-CGARCH respectively are 0.738 and .483 and the mean absolute error of them are 0.510  and .3804.}
\begin{table}
\vspace{-.05in}
\caption{{\small Descriptive statistics  for the  simulated data (sample size=300).}}
\hspace{1.3cm}
\begin{tabular}{c c c c c c}
  \hline
   Mean & Std. dev. & Skewness & Maximum & Minimum &  Kurtosis  \\

  \hline
  0.034&  0.860& 0.289& 3.109 &-2.997  &4.502 \\

   \hline
  \end{tabular}
\end{table}
\textbf{\section{Empirical Applications}
We apply the  daily stock market index of  Dow Jones industrial average (DJIA) from 07/10/2009 to 14/12/2010 (300 observations) and $S\&P500$ from 12/12/2006 to 22/02/2008 (300 observations)  for estimation. Figures 3  demonstrates the stock market index and the percentage returns \footnote{Percentage returns are defined as $r_t=100*\log(\frac{P_t}{P_{t-1}})$, where $P_t$ is the index level at time t.} of both DJIA and $S\&P500$. It is evident  that the stock market index of DJIA and $S\&P500$ have the divers of shocks. A summary of descriptive statistics of these returns are in Table 3.}\\

\begin{table}
\vspace{-.05in}
\caption{{\small Results of the Bayesian Estimation of the  simulated MS-CGARCH model. }}
\hspace{4cm}
\begin{tabular}{c c c c }
  \hline

    & True values &  Mean & Std. dev.  \\
    \hline
   $a_{01}$ &2.200  &2.301  & 0.415 \\
   $a_{11}$ &0.750 &0.721  & 0.060 \\
   $a_{21}$ &0.150  &0.147  & 0.047\\
   $b_{01}$ &0.700  &0.661  & 0.085  \\
   $b_{11}$ &0.300  &0.270  & 0.070 \\
  $b_{21}$ &0.200  &0.213 & 0.056\\
  $a_{02}$ & 0.400 &0.361  &0.084 \\
  $a_{12}$ &0.150  &0.176  & 0.043 \\
    $a_{22}$ &0.100 &0.119  & 0.056 \\
  $b_{02}$ & 0.200 &0.181 &0.094  \\
  $b_{12}$ &0.100  &0.050  & 0.026 \\
   $b_{22}$ &0.200  &0.203  &0.081\\
  $\gamma_1$ &2.000 &2.01  & 0.603 \\
  $\gamma_2$ &0.500 &0.742 & 0.150 \\
  $\eta_{11}$ &0.850&0.620 & 0.086 \\
  $\eta_{22}$ &0.950  &0.869  &0.042  \\
  \hline

\end{tabular}
\end{table}
\par \textbf{In Tables 4 and 5, the posterior means and standard deviations from the estimation of MS-CGARCH and MS-GARCH models for DJIA  and $S\&P500$ daily returns are reported.  The results of estimating MS-GARCH in both cases (DJIA and $S\&P500$ daily returns) show that the first regime is the high volatility regime. In the high volatility state, the conditional variance is more sensitive to recent shocks ($a_{11}>a_{21}$) and less persistence ($a_{21}<a_{22}$) than the low volatility regime. Also the outcomes of estimating MS-CGARCH (Tables 4 and 5) show that the first regime is the high volatility state that in each state the first  component is higher volatile than the second one. The values of $\gamma_1$ and $\gamma_2$ show the speed of transition (in each regime) from one component to the other one. This specification causes  the MS-CGARCH to be more flexible than the MS-GARCH to capture the variants of shocks: very high, high, moderate and low shocks. Indeed the MS-CGARCH model is able to model the gradual changes in high  and low  volatile states by the effect of their components in each state.
\\
 We  compare the forecasting volatility of each model with the squared returns. Figure 4 and  5  show that the forecasting volatility of MS-CGARCH is much better than MS-GARCH model. Table 6  reports the measures  of performance forecasting, the   mean absolute error and root of mean squared error, for both MS-CGARCH and MS-GARCH models. Based on the results given in Table 6, the MS-CGARCH model has a much  better forecast than MS-GARCH model.}


\begin{table}
\vspace{-.05in}
\caption{{\small Descriptive statistics  of DJIA and $S\&P500$ stock market returns.}}
\hspace{.2cm}
\begin{tabular}{c c c c c c c}
  \hline
  & Mean & Std. dev. & Skewness & Maximum & Minimum &  Kurtosis  \\

  \hline
  DJIA&0.05& 1.01& -0.18& 3.82 &-3.67  &4.76  \\

$S\&P500$&-0.01 &1.04 &-0.54 &2.87 &-3.53 &4.14\\
   \hline
  \end{tabular}
\end{table}

\label{fig3}
\input{epsf}
\epsfysize=5in
 \begin{figure}
\centerline{\epsffile{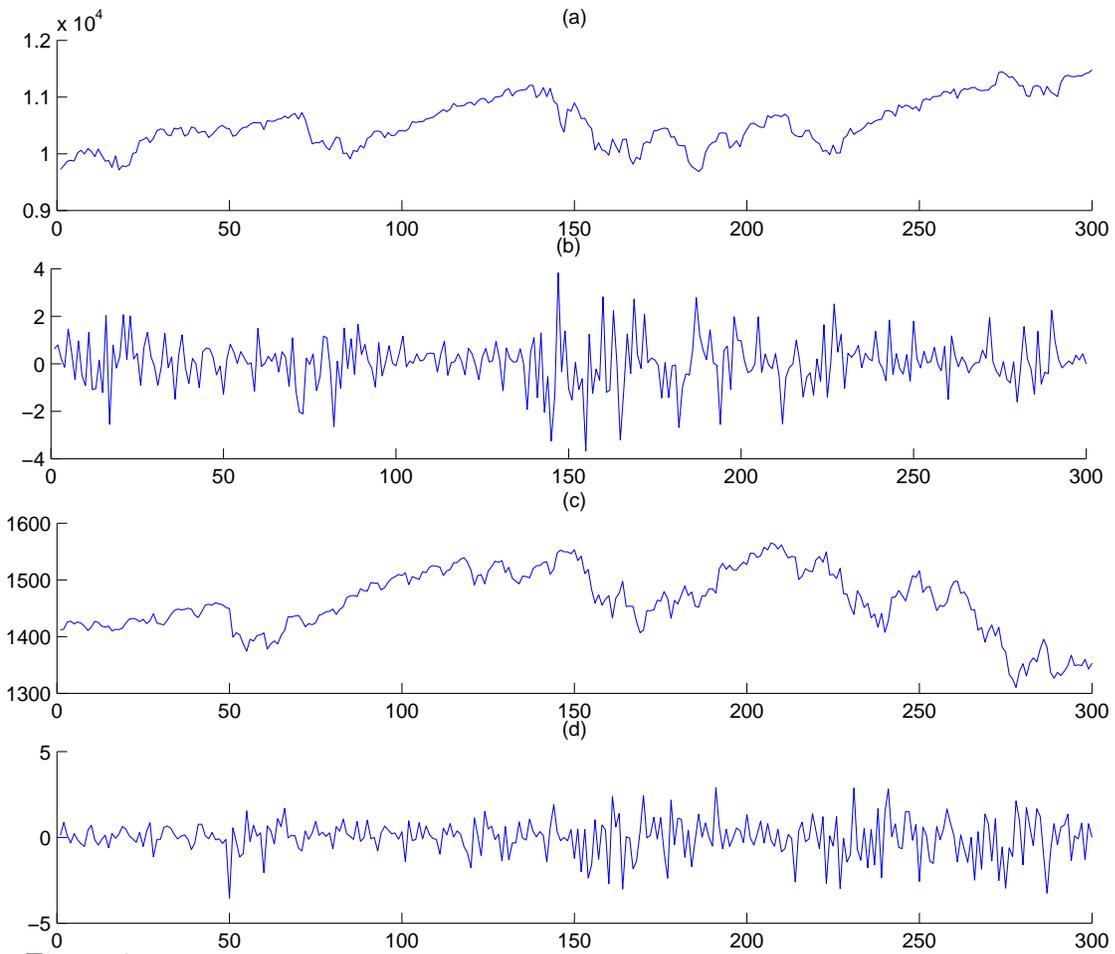}}
\vspace{-.2in}
\caption{\scriptsize (a): DJIA stock market index, (b): Percentage daily returns of DJIA, (c):  $S\&P500$   stock market index and (d): Percentage daily returns of $S\&P500$.}
\end{figure}
\label{fig4}
\input{epsf}
\epsfysize=3.3in
 \begin{figure}
\centerline{\epsffile{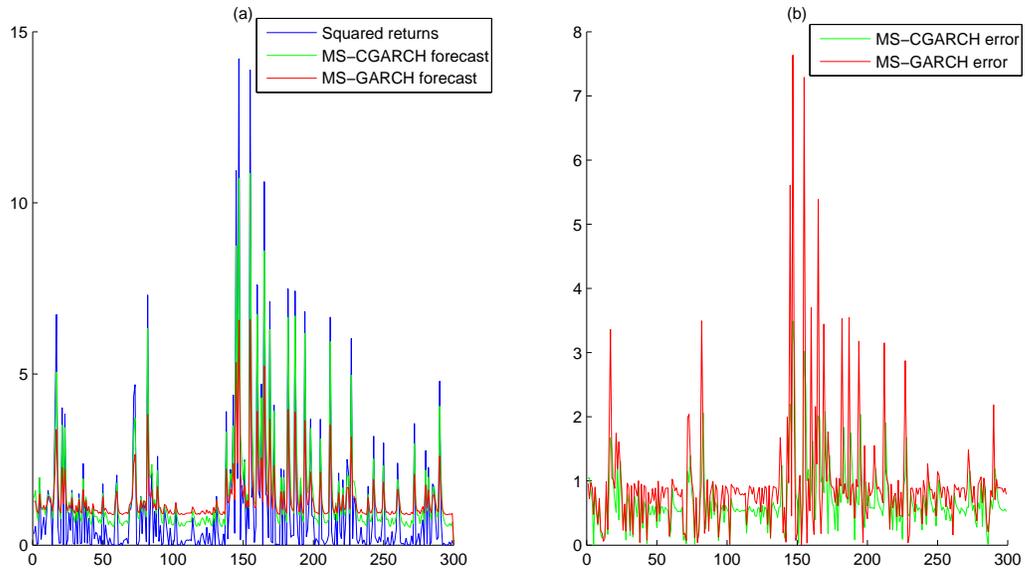}}
\vspace{-.2in}
\caption{\scriptsize (a): Squared returns of  DJIA  (blue), forecast by MS-GARCH (red) and forecast by MS-CGARCH (green). (b): Absolute forecast error of squared returns (DJIA)  in the MS-GARCH (red) and in the MS-CGARCH (green).}
\end{figure}
\begin{table}
\vspace{-.05in}
\caption{{\small Posterior means and standard deviations (DJIA daily returns). }}
\hspace{3cm}
\begin{tabular}{lllllll}
\hline
 & \multicolumn{2}{c}{}MS-CGARCH & & & \multicolumn{2}{c}{}MS-GARCH \\
\cline{2-3}\cline{6-7}
 &Mean  &Std.dev.  & & &Mean &Std.dev  \\
\hline
 $a_{01}$& 3.150 & 0.590 & &  &1.859  &0.316  \\
 $a_{11}$& 0.651 &0.145  &  & & 0.504 & 0.102 \\
 $a_{21}$& 0.094 & 0.043 &  & & 0.216 & 0.054 \\
 $b_{01}$& 0.821 & 0.142 &  & &$\_$ & $\_$ \\
 $b_{11}$& 0.306 & 0.061 &  & &$\_$ & $\_$ \\
 $b_{21}$& 0.277 & 0.050 &  & &$\_$& $\_$ \\
 $a_{02}$&0.658  & 0.157 &  & & 0.498& 0.091 \\
 $a_{12}$& 0.296 & 0.054 &  & &0.189 & 0.055 \\
 $a_{22}$& 0.203 & 0.053 &  & & 0.242& 0.071 \\
 $b_{02}$&0.291  & 0.049 &  & &$\_$ & $\_$ \\
 $b_{12}$&0.092  & 0.049 &  & &$\_$ & $\_$ \\
 $b_{22}$&0.334  &0.087  &  & &$\_$ & $\_$ \\
 $\gamma_1$&1.554  & 0.307 &  & &  $\_$& $\_$ \\
 $\gamma_2$& 0.756 & 0.136 & & & $\_$ & $\_$ \\
 $\eta_{11}$& 0.806 & 0.140 & & &0.542  & 0.092 \\
 $\eta_{22}$& 0.941 & 0.038 & & & 0.899 & 0.031 \\
 \hline
\end{tabular}
\end{table}
\label{fig5}
\input{epsf}
\epsfysize=3.3in
 \begin{figure}
\centerline{\epsffile{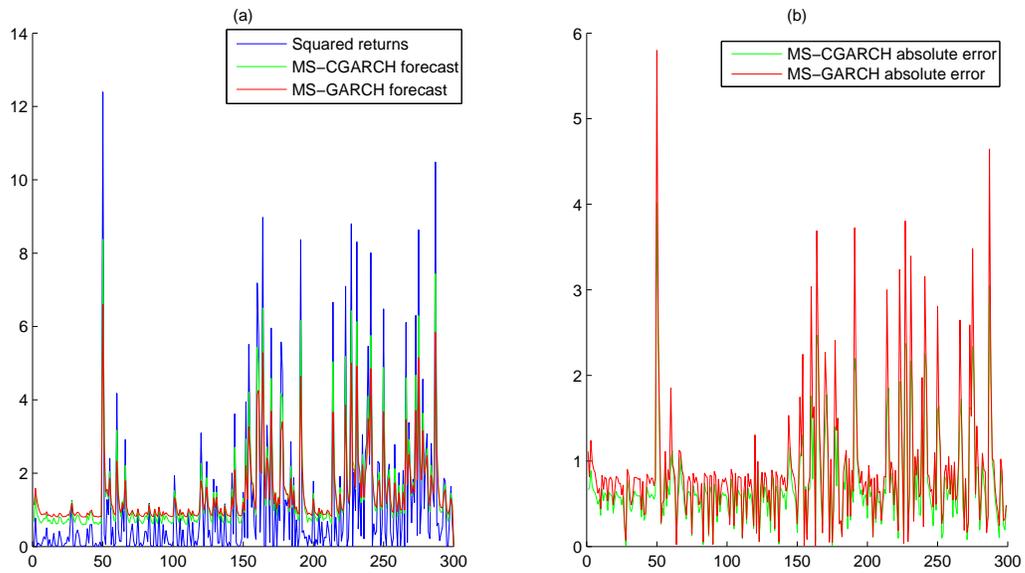}}
\vspace{-.2in}
\caption{\scriptsize (a): Squared returns of  $S\&P500$  (blue), forecast by MS-GARCH (red) and forecast by MS-CGARCH (green). (b): Absolute forecast error of squared returns ($S\&P500$)  in the MS-GARCH (red) and in the MS-CGARCH (green).}
\end{figure}
\begin{table}
\vspace{-.05in}
\caption{{\small Posterior means and standard deviations ($S\&P500$ daily returns). }}
\hspace{3cm}
\begin{tabular}{lllllll}
\hline
 & \multicolumn{2}{c}{}MS-CGARCH & & & \multicolumn{2}{c}{}MS-GARCH \\
\cline{2-3}\cline{6-7}
 &Mean  &Std.dev.  & & &Mean &Std.dev  \\
\hline
 $a_{01}$& 2.016 &0.548  & &  & 1.330 &0.329  \\
 $a_{11}$& 0.609 & 0.132 &  & &  0.454& 0.102 \\
 $a_{21}$& 0.184 & 0.050 &  & &0.303  & 0.062 \\
 $b_{01}$& 0.767 & 0.129 &  & &$\_$ & $\_$ \\
 $b_{11}$& 0.286 & 0.053 &  & &$\_$& $\_$ \\
 $b_{21}$& 0.352 & 0.059 &  & &$\_$ & $\_$ \\
 $a_{02}$& 0.622 & 0.145 &  & &0.500 & 0.092 \\
 $a_{12}$& 0.249 & 0.092 &  & &0.162 & 0.069 \\
 $a_{22}$& 0.130 & 0.056 &  & &0.232 & 0.065 \\
 $b_{02}$& 0.313 &0.058  &  & & $\_$& $\_$ \\
 $b_{12}$& 0.086 & 0.049 &  & & $\_$& $\_$ \\
 $b_{22}$& 0.315 & 0.077 &  & &$\_$ & $\_$ \\
 $\gamma_1$& 1.856 & 0.505 &  & & $\_$ &$\_$  \\
 $\gamma_2$&0.725  &0.126  & & &  $\_$& $\_$ \\
 $\eta_{11}$& 0.774 &0.086  & & &0.821  & 0.065 \\
 $\eta_{22}$&0.915  &0.027  & & & 0.936 & 0.022 \\
 \hline
\end{tabular}
\end{table}
\begin{table}
\vspace{-.05in}
\caption{{\small Measures of performance forecasting. }}
\hspace{.5cm}
\begin{tabular}{ c c c c c c}
  \hline

    &\multicolumn{2}{c}{DJIA}&& \multicolumn{2}{c}{$S\&P500$} \\
   \cline{2-3} \cline{5-6}
     &  MS-GARCH & MS-CGARCH && MS-GARCH & MS-CGARCH  \\
    \hline
   RMSE & 1.281 & 0.834 & &1.169& 0.902  \\
    MAE& 0.940 & 0.687 & &0.904&0.723  \\

  \hline

\end{tabular}
\end{table}

\section{Conclusion}
In this  paper  a generalization  of the MS-GARCH model  has been  presented where the conditional variance in each state is a convex combination of two different \textbf{GARCH} components with time varying coefficients, one of the component with higher volatility than the other component. \textbf{The structure of the model makes a dynamic behavior in each regime to react differently to the species of shocks.} Our model can provide more better forecast of volatility toward MS-GARCH model.    For the estimation of parameters we have applied the Bayesian estimation algorithm.  We provide a simple necessary and sufficient condition for the existence of an upper bound for the second moment.
 \par This work has the potential to be applied in the context of financial time series. The empirical distribution of daily returns doesn't generally have a Gaussian distribution.
They have fat tails densities (they are called leptokurtic). One of  the extending of this work is  considering the  fat tail densities instead of Gaussian distribution, that can  cause better modeling of the financial time series. \textbf{Also we can generalize this model by allowing an  ARMA structure for the conditional mean.} \\

\bibliographystyle{plain}

\end{document}